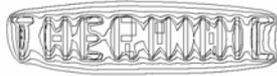

# THE HOT-SPOT PHENOMENON AND ITS COUNTERMEASURES IN BIPOLAR POWER TRANSISTORS BY ANALYTICAL ELECTRO-THERMAL SIMULATION

*Fabio Stefani, Paolo Emilio Bagnoli*

Dept. of Information Engineering, University of Pisa, via Caruso 16, I-56122 Pisa, Italy
Ph +39 0502217503, Fax +39 050 2217522, Email: fabio.stefani@iet.unipi.it, p.bagnoli@iet.unipi.it

**ABSTRACT**

This communication deals with a theoretical study of the hot spot onset (HSO) in cellular bipolar power transistors. This well-known phenomenon consists of a current crowding within few cells occurring for high power conditions, which significantly decreases the forward safe operating area (FSOA) of the device. The study was performed on a virtual sample by means of a fast, fully analytical electro-thermal simulator operating in the steady state regime and under the condition of imposed input base current. The purpose was to study the dependence of the phenomenon on several thermal and geometrical factors and to test suitable countermeasures able to impinge this phenomenon at higher biases or to completely eliminate it. The power threshold of HSO and its localization within the silicon die were observed as a function of the electrical bias conditions as for instance the collector voltage, the equivalent thermal resistance of the assembling structure underlying the silicon die, the value of the ballasting resistances purposely added in the emitter metal interconnections and the thickness of the copper heat spreader placed on the die top just to the aim of making more uniform the temperature of the silicon surface.

## 1. INTRODUCTION

The well-known "hot spot" is a phenomenon occurring in power devices, bipolar transistors [1,2] working in high conduction regimes and also MOS [3], and consists of a particular situation in which, due to the interaction among geometrical, electrical and thermal factors, a small part of the device accumulates most of the total current. This causes, of course, a local dramatic temperature uprising that can, but not necessarily, provoke a failure of the device. This phenomenon, whether damages the transistor or not, is anyway dangerous for the device reliability and causes a serious limitation for the forward safe operating area (FSOA) of a device and its maximum power ratings. Although the reasons of the hot spot onset were deeply investigated [2] and are well known, reliable and suitable electro-thermal simulation tools may be widely useful in order to explore all the conditions in which it occurs (HSO) and to test technological solutions able to retarding or inhibit it. The electro-thermal simulation procedure used in the present work, being based on fully analytical models and in the steady state regime, differs from the usual strategies involving the dynamic behavior of the device and finite elements calculation tools [1].

The present simulation method, elsewhere presented [4,5] applied to a real power device, were experimentally validated in terms of prediction of HSO base current and localization within the silicon die [5]. Thanks to this agreement between the simulations and the experiments and to the relatively high calculation speed offered by this method, an extensive theoretical study of the HSO was performed on a virtual sample as a function of several but not complete thermal, electrical and device layout characteristics. Furthermore the effectiveness of two traditional countermeasures, the ballasting emitter resistance at the metal fingers input and a top copper heat spreader, were explored as a function of the structural parameters, as for instance the copper layer thickness and the thermal properties of the copper-silicon interface.

## 2. THE DEVICE UNDER TEST

The layout of the device used for the present study is shown in Fig. 1a. It is a bipolar power transistor with cellular structure in which the emitter cells are organized in a regular matrix with 30 columns and 11 rows of square cells, 50 microns wide. The die size is 3.25x1.35 $mm^2$, 300 microns thick. The collector contact is placed on the bottom side of the die. Because of the adiabatic boundary conditions on the top and lateral sides, we should consider this structure as the half of the whole symmetrical device. Fig. 1a also shows the interdigitated structure of the metal interconnections and the positions of the current pads for both the emitter and the base. A single column of emitter cells is connected with the emitter distribution metal by a single finger. On the other side the base contacts of the same column are similarly connected by means of a base metal finger.

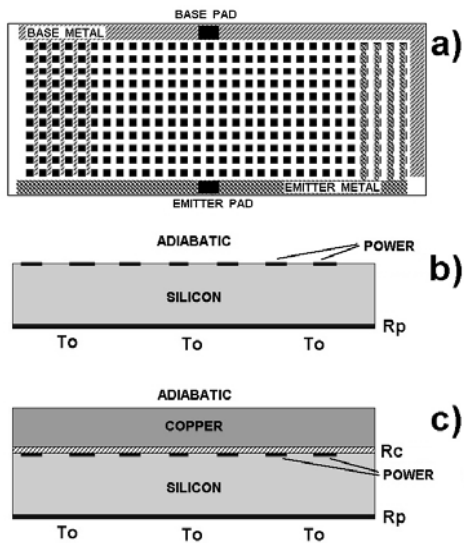

*FIG. 1: a) Layout of the device used for the simulation. Device cross-sections without (b) and with (c) a top heat spreader.*

The figures 1b and 1c show the lateral view of the two structures used for the thermal simulation: a standard one-layer structure (Fig 1b) in which the thermal contribution of the package is represented by a specific contact thermal resistance Rp placed between the bottom die surface and the heat sink at uniform temperature To. The other structure (Fig.1c) was designed to study the effectiveness of a top highly conducting metal layer (copper) in retarding or eliminating the hot spot onset. The power generating cells are still located on the top silicon but they are insulated from the top copper spreader by means of a thin insulator layer, whose thickness and thermal conductivity are summarized in an interface contact thermal resistance per unit area Rc. The thickness of the metal layer and the Rc values are parameters under investigation in the present work.

### 3. SIMULATION PROCEDURE

The whole simulation procedure is based on the research of the device electro-thermal steady-state condition under given electrical (collector voltage and total base current bias) and thermal (device assembling structure, heat transfer boundary conditions, package thermal resistance) input data. This procedure is composed by the cyclic interaction between two fast analytical electrical and thermal solvers, as schematically shown in the diagram of Fig. 2. The first one calculates the electrical power distribution within the emitter cells starting from their own temperatures and the total input base current. The second one operates in the reverse direction: from the power dissipation in the cells it can give the exact temperature distribution according to the device assembling structure and boundary conditions.

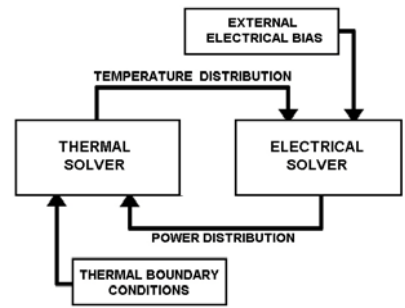

*FIG. 2: Flow diagram of the simulation procedure.*

The electro-thermal steady state equilibrium can be reached - i.e. the iterative procedure converges - both below and above the hot spot onset just because of two reasons:
a) the constant input base current condition, who has a stabilizing effect on the bias point avoiding the thermal runaway of the transistor, that is instead activated by the constant base voltage condition;
b) the decreasing behaviour of current gain for high injection conditions (see Fig. 3b) and the inversion of the current gain temperature dependence, from positive to negative.
As a consequence in the hot spot condition the total power and the average die temperature undergo a dramatic decrease determined by the lessening of the transistor static current gain [6]. The cyclic procedure is halted when the total electrical power variation goes under a given accuracy. It was found that it quickly converges for base currents below the HSO, less quickly converges for base currents above HSO and it diverges in presence of thermal runaway.

### 3.1. Thermal solver

The thermal solver, whose aim is the calculation of the top silicon temperature distribution, is based on the analytical thermal simulation system DJOSER, elsewhere described [7] and specifically designed for multi-layer stepped pyramidal structures. This simulator, in which the power generation is two-dimensional and located at the horizontal surfaces of any slab, is able to take into account also the presence of contact thermal resistances per unit area between two slabs, representing thin soldering or attaching layers. The use of the DJOSER thermal simulator allows a fine and realistic description of the device package whose thermal resistance play a mayor role in the determination of the hot spot characteristics. Beside this, for practical purposes in the present work, minimalist device thermal structures were used, assigning to the bottom contact thermal resistance Rp the role of representing the whole package thermal contribution.

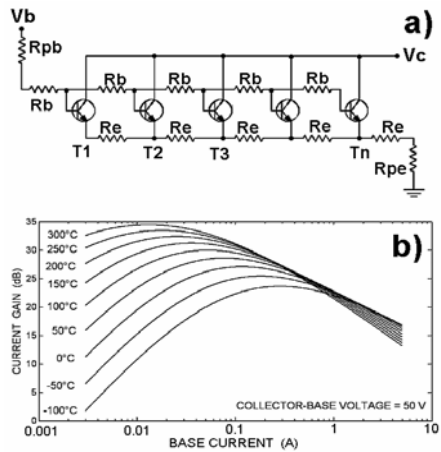

FIG. 3: a) Equivalent circuit of a single column of cells. b) Current gain characteristics of the device.

In the case of the single layer structure (Fig. 1b), the emitter cell temperatures were calculated using a direct explicit analytical expression which consists of a cosine-like double Green series using infinite set of eigenvalues for the x and y directions. However the series were truncated at a given maximum value which is a best compromise between temperature accuracy and calculation time. The whole power generation distribution in all the emitter cells is directly involved in the analytical expression of the single cell temperature. In the double-layer structure (Fig. 1C) the cell temperatures were calculated from the resolution of an algebraic system having the temperatures and the heat flux values in the centers of the cells as unknown variables. The matrix of the system coefficients, still composed by the double harmonic series, was separately calculated once a time and memorized, since the power generation density, which changes at every iteration step, is involved in the know term vector of the algebraic system only. This last procedure allowed us to largely reducing the calculation time of the thermal solver, speeding up the iterative process for this type of structure too.

### 3.2. Electrical solver

Fig. 3a shows the equivalent electrical circuit for a single column of cells represented by the bipolar transistors. The resistances Re and Rb are referred to the metal vertical distances, for emitter and base metals respectively, between two adjacent cells and are the same for all the columns. Rpe and Rpb, also addressed as 'penalty' resistances, which are different for the various columns, are the resistances due to the distribution wider metal stripes of emitter and base respectively. More precisely, the penalty resistances of each column are calculated on the basis of the distance from the corresponding pad and on the current values (obtained in a previous iteration step) flowing in the columns between the actual one and the pad, so that each metal finger may be considered as connected to the pad with the proper real voltage drop between the pad and the finger input.

The transistors were modeled by means of the well-known relationships holding for a single cell:

$$I'_E = I_S(\exp(V_{BE}/V_T) - 1) \qquad I'_B = I'_E /(h_{FE} + 1) \qquad (1)$$

where $I'_E$ and $I'_B$ are the single cell currents, $I_S$ is the reverse saturation current of the base-emitter junction, $V_T$ is the thermal voltage, $\eta$ (set at 1.6) is the ideality factor of the base-emitter junction and $h_{FE}$ is the static current gain. As a first approximation and due to the relatively high silicon thickness, the electrical power was supposed to be localized on the silicon surface within the emitter cells. The temperature dependence of every electrical parameter was carefully modeled. For $h_{FE}$ the dependence on temperature, base current and collector-base voltage has been derived from previous experimental results [1,4,5] carried out under pulsed regime and isothermal conditions, obtaining an empirical $h_{FE}(T, I'_B, V_{CB})$ relationship shown in Fig 3b. The usual temperature dependence - i.e. doubling every 10K of temperature increment - has been used for $I_S$. All the transistor parameters were evaluated with the respective transistor temperature given by the thermal solver, while the network resistors Rb and Re were supposed to depend on the averaged finger temperature only. Under these simplifications and assumptions, the electrical problem was reduced to the resolution of a number of circuits equal to the number of columns of cells. Since the device was supposed to be driven with a constant base current Ib, we had to use a double iterative method. Firstly, every finger was solved with a constant base and collector voltage, then the resulting base currents were summed, thus obtaining the total current. The imposed base voltage was adjusted and all the finger circuits were recalculated until the total of currents equates the value of Ib. This outer cycle employs a second order numerical algorithm and usually converges in a few iterations. The inner cycle, which is carried out for all the fingers, is based by the alternate resolution of the resistor network from external voltages and transistor currents to obtain the intrinsic transistor-cell voltages and their currents from equations (1) and therefore the power dissipated within the cells.

### 4. SIMULATION RESULTS

### 4.1. Hot spot characterization.

All the simulation tests were carried out by scanning the input base current Ib from 0.5 mA to a value well above the HSO condition and using a very small current step (0.1 mA). For each Ib value, the power distribution within the emitter cells was memorized so that the corresponding temperature map can be calculated at any time with any spatial resolution by means of the thermal solver.

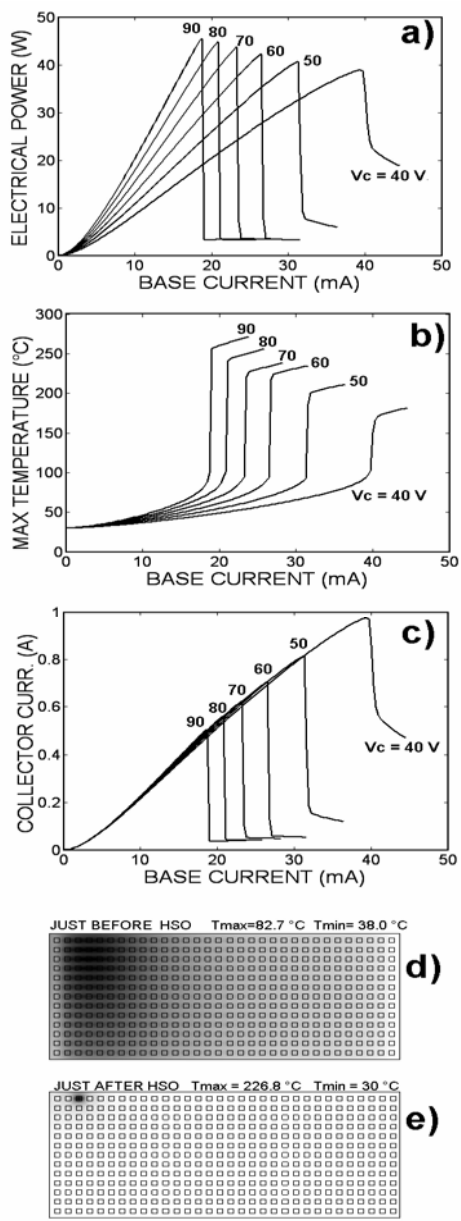

*FIG. 4: Plots of electrical power (a), maximum temperature (b) and collector current (c) versus base current for several collector voltage values. Gray-scaled temperature maps just before (d) and after (e) the hot spot, for Vc =60 V.*

Fig. 4 shows the typical behavior of the temperature and the main electrical parameters of a device affected by the hot spot. The tests related to the figure were carried out placing the emitter pad on the left limit of the distribution metal and the base one on the right limit. Furthermore the bottom specific thermal resistance Rp was set to 0.5

mm$^2$°C/W and the heat sink temperature was 30 °C. The three plots traced as a function of the input base current were calculated for several values of the collector voltage. As can be seen, as the HSO abruptly occurs, it causes a dramatic fall of the electrical power (Fig. 4a) and the collector current (Fig. 4c) which corresponds to an abrupt arising – generally more than one hundred degrees - of the maximum temperature on the die (Fig. 4b). For biases above the HSO condition, the values of all these parameters reach an equilibrium condition which only weakly depends on bias. It should be noted the different ways in which the HSO phenomenon appears for different collector voltages. For low Vc values the abrupt variations of the power, of Ic and of the maximum temperature seem to be lesser than for higher voltages. This is consistent with the fact that a particular collector voltage value exists below which the hot spot cannot be activated for any current biases.

Fig 4d and 4e show the temperature maps for Vc = 60V just before (d) and after (e) the HSO drawn in gray scale. As can be seen, the birth site of the phenomenon is on the same abscissa of the emitter pad (left) - at the cell column having the lesser emitter penalty resistance - and near the base distribution metal. This seems to demonstrate that the emitter metal plays an important role in setting the hot spot localization site, greater than that of the base metal.

### 4.2. Influence of package thermal resistance.

It is quite obvious that the package thermal resistance, in the present case represented by the bottom contact specific thermal resistance Rp, is a serious source of degradation for the power device performances. In order to quantitatively study this effect on the HSO characteristics, several simulation tests were carried out as a function of Rp in the range 0-70 mm$^2$°C/W, which corresponds to the range 0-16 °C/W of package thermal resistance for the present die size. Fig. 5a shows the hyperbolic-like decreasing plot of the HSO power versus Rp in both the scales. From the obtained results it is evident that the presence of a package thermal resistance causes two main effects. At first the threshold power for the hot spot onset is a strongly decreasing function of Rp, and most of this decrease occurs just in the range of very low values of Rp. This clearly emphasizes the dramatic importance of package thermal design from the heat dissipation point of view in avoiding such unsuitable effects and in keeping the maximum operating power as high as possible. This care must be kept not only by the device producer but also by the device customer user since the thermal resistance is due to the whole assembling structure, including the package, built by both these industrial subjects. The second effect concerns the migration of the hot spot onset point.

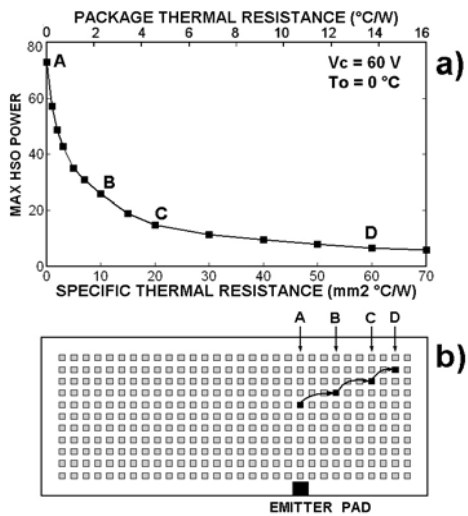

*FIG. 5: a) Plot of HSO power vs package thermal resistance Rp for Vc=60V. b) Hot spot site migration path with increasing Rp.*

In the Fig. 5b the localization of the phenomenon is traced for four values of Rp, also marked in the upper plot. It is evident that for low thermal resistance values the temperature peak is practically localized in the same point of the case Rp=0 (i.e. at the abscissa of the emitter pad, in this case placed at 2/3 of the structure), but, as Rp increases, it progressively migrates toward the corner of the chip area occupied by the cell grid. This behavior practically means that for low thermal resistances the electrical factors tend to prevail on the thermal ones in determining the HSO birth site, while the contrary happens for high values of Rp. In fact the corner of the die is quite disadvantaged for the heat dissipation with respect of the central part of the surface due to the adiabatic boundary conditions at the lateral surfaces of the chip. An adiabatic surface practically acts on a given heat source as a mirror and can be modeled with the presence of an imaging source placed at the same but opposite distance: as closer the heat source is to the adiabatic boundary, as much its imaging counterpart heats it and its effective temperature is higher. Therefore the balancing between electrical and thermal effects produces the peak progressive migration on the die surface.

As a final consideration and as can be seen in the next sections, high values of the bottom thermal resistance inhibit or decrease the benefits on the HSO characteristics offered by suitable countermeasures.

**4.3. Countermeasure: the emitter ballasting resistor.**

The role of the emitter ballasting resistors Reb in preventing thermal instabilities such as second breakdown or hot spot formation was deeply investigated [8]. In fact, a small polysilicon resistance at the output of each emitter metal finger introduces a negative feedback subtracting biasing to the finger devices and helping to stabilize them. This favourable property is punctually confirmed by the results in the present work.

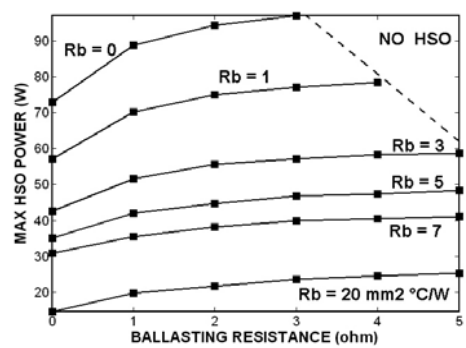

*FIG. 6: HSO power versus ballasting resistance Reb for several bottom thermal resistances Rp:*

Fig 6 shows the maximum power at the HSO as a function of the ballasting resistance Reb, in the range 1-5 ohm and for several values of the package specific thermal resistance Rb, from 0 to 20 mm$^2$ °C/W. As can be seen the relative power increase due to the presence of the ballast is up to about 37 % for all the Rb values. Furthermore, the upper right corner of the graph in Fig.6, the area limited by the dashed line indicates the sets of values (Rb< 3 mm$^2$ °C/W, Reb > 3 ohm) for which the hot spot does not occur. In the above simulations the power dissipation due to the joule heating in the ballasting resistors was neglected since in the range of Reb used, it is well negligible with respect to the total power handled by the device. However the increase of the ballasting resistance also implies an increase of the power dissipation within these resistances and a possible degradation of the device electrical characteristics may occur.

**4.4. Countermeasure: the top metal heat spreader.**

The presence of a high thermally conductive layer above the power dissipating silicon surface, with adiabatic top surface too, may contribute to spreading off the generated heat and making the temperature of silicon more uniform. Therefore this structural detail may act as a hot spot countermeasure, retarding or eliminating this phenomenon. However the effectiveness of its action depends on several factors as the metal thermal conductivity, the layer thickness and overall the thermal conducting properties of the thin layer which insulates the power emitter cells and their interconnections from the metal layer. Beside silver has a greater thermal conductivity value than copper, for the test simulations we used this last material for obvious cost reasons from the application point of view. These simulations were carried out using the two-layers thermal solver. Fig. 7 shows the simulation results for this type of countermeasure; in Fig. 7a the power vs base currents plots are shown for several values of the copper layer thickness.

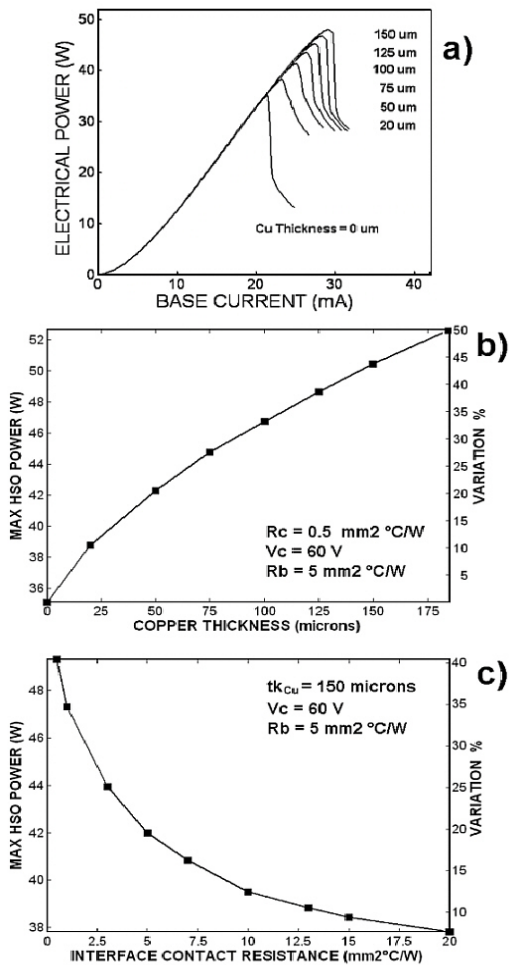

FIG. 7: a) Electrical power vs base current for various copper thickness. Plots of HSO power and percentage variations versus copper thickness.(b) and versus the interface thermal resistance Rc (c).

The maximum power occurring at HSO is plotted in Fig. 7b in terms of watt (left vertical scale) and percentage increase (right vertical scale) with respect of the zero-thick copper. The simulation conditions, in term of collector voltage (Vc), bottom (Rb) and silicon-copper interface (Rc) specific thermal resistances are reported within the figure. As can be seen the maximum power before the HSO, and therefore the maximum base and collector currents, are increasing functions of the copper thickness due to the redistribution of the heat flux across the silicon surface.

On the other hand, the insulating properties of the copper-silicon interface works as a negative factor for the HSO inhibition. In fact, as the specific thermal resistance increases, the advantages of the top metal layer are greatly reduced, as it is demonstrated by the plot of Fig.

7c showing in watt and in percentage increase, for a copper thickness of 150 microns only, the decreasing behavior of the max HSO power as a function of the interface contact thermal resistance.

These simulation tests quantitatively demonstrated that this structural detail may be very useful to impinge the HSO at higher bias conditions. However it is straightforward to imagine that better improvements can arise with the presence of convective boundary conditions at the copper layer top surface or with the use of an up-down flipped structure, in which the silicon die is mounted in a flip-chip-like configuration and the wider copper layer is thermally connected to the package and the bottom heat sink. However, the simulations concerning these last configurations are still to be performed.